\documentclass[runningheads]{llncs}

\usepackage[a-1b]{pdfx}

\usepackage[most]{tcolorbox}
\newtcolorbox{bluetapbox}[1]{
	fontupper=\footnotesize,
	enhanced,
	title={#1},
	colback=blue!3,
	colframe=blue!40,
	colbacktitle=blue!50!black,
	coltitle=white,
	fonttitle=\bfseries,
	boxrule=0.2pt,
	arc=2mm,
	left=2mm,
	right=2mm,
	top=0mm,
	bottom=0mm,
}

\newtcolorbox{greentapbox}[1]{
	fontupper=\footnotesize,
	enhanced,
	title={#1},
	colback=green!3,
	colframe=green!40,
	colbacktitle=green!50!black,
	coltitle=white,
	fonttitle=\bfseries,
	boxrule=0.2pt,
	arc=2mm,
	left=2mm,
	right=2mm,
	top=0mm,
	bottom=0mm,
}

\usepackage{empheq}
\usepackage{fancybox}
\usepackage{amsmath, amsfonts}
\usepackage{amssymb}
\usepackage{graphicx}
\usepackage{placeins}
\usepackage{textcomp}
\usepackage{xcolor}
\usepackage{subcaption}
\usepackage{xspace}
\usepackage{tikz} 
\usepackage{bbding} 
\usepackage{diagbox} 
\usepackage[normalem]{ulem}

\usepackage{algorithm}
\usepackage{algpseudocodex}
\usepackage{algorithmicx}

\usepackage{environ}
\usepackage{adjustbox}

\usepackage{pgfplots}
\usetikzlibrary{patterns}
\usepgfplotslibrary{groupplots}

\usepackage{todonotes}

\usepackage{enumitem}

\usepackage[most]{tcolorbox}

\usepackage{wrapfig}

\usetikzlibrary{spy}

\usepackage{tabularx}
\usepackage{soul}

\usepackage{multirow}

\usepackage{capt-of}

\usepackage{listings}
\definecolor{backcolour}{rgb}{0.95,0.95,0.94}
\lstdefinelanguage{Rocq}{
    keywords=[1]{Definition, Record, Inductive, Parameter, Axiom, Theorem, Lemma, Proof, Qed, Defined, Prop, Type, forall, exists, fun, match, with, end, let, in, where},
    keywords=[2]{relation, Ensemble, clos_trans, transitive},
    sensitive=true,
    morecomment=[s]{(*}{*)}
}
\lstdefinestyle{coqstyle}{
    backgroundcolor=\color{backcolour},
    commentstyle=\color{olive}\rmfamily\itshape,
    keywordstyle=[1]\color{blue}\bfseries,
    keywordstyle=[2]\color{violet}\bfseries,
    keywordstyle=[3]\bfseries,
    numberstyle=\scriptsize\color{gray},
    basicstyle=\linespread{0.95}\ttfamily\small,
    basewidth=0.5em,
    breakatwhitespace=false,
    breaklines=true,
    captionpos=b,
    columns=fixed,
    fontadjust=true,
    frame=single,
    keepspaces=true,
    mathescape,
    numbers=left,
    numbersep=5pt,
    rulecolor=\color{backcolour},
    showspaces=false,
    showstringspaces=false,
    showtabs=false,
    tabsize=2
}
\usepackage{mathpartir}
\usepackage{mathtools}

\usepackage{marginnote}

\usepackage{balance}

\usepackage{booktabs}

\newcommand{\inlsec}[1]{\smallskip\noindent\textbf{#1.}}

\newcommand{\ci}{{Cut Isolation}\xspace}
\newcommand{\rc}{{Read Committed}\xspace}
\newcommand{\ra}{{Read Atomicity}\xspace}
\newcommand{\tcc}{{Transactional Causal Consistency}\xspace}
\newcommand{\si}{{Snapshot Isolation}\xspace}
\newcommand{\ser}{{Serializability}\xspace}

\definecolor{ps1green}{HTML}{B7E6A5} 
\definecolor{ps2green}{HTML}{7CCBA2}
\definecolor{ps3green}{HTML}{46AEA0}
\definecolor{ps4green}{HTML}{089099}
\definecolor{ps5green}{HTML}{00718B}
\definecolor{ps6green}{HTML}{045275}
\definecolor{ps7green}{HTML}{003147}

\definecolor{diverge1}{HTML}{045275}
\definecolor{diverge2}{HTML}{089099}
\definecolor{diverge3}{HTML}{EF8733}
\definecolor{diverge4}{HTML}{FCDE9C}
\definecolor{diverge5}{HTML}{F0746E}
\definecolor{diverge6}{HTML}{DC3977}
\definecolor{diverge7}{HTML}{7C1D6F}

\definecolor{equalsoneoptcolor}{RGB}{253, 243, 208}
\definecolor{bigandsmallcyclesoptcolor}{RGB}{217, 232, 214}
\definecolor{otheroptcolor}{RGB}{251, 231, 207}

\newcommand{\set}[1]{\{#1\}}

\newcommand{\Key}{{\sf Key}}

\newcommand{\Val}{{\sf Val}}

\newcommand{\h}{\mathcal{H}}

\newcommand{\readevent}{{\sf R}}
\newcommand{\Read}{{\sf Reads}\;}
\newcommand{\ReadTx}{{\sf ReadTx}}
\newcommand{\writeevent}{{\sf W}}
\newcommand{\Write}{{\sf Writes}\;}
\newcommand{\WriteTx}{{\sf WriteTx}}

\newcommand{\SO}{\textsf{\textup{SO}}}
\newcommand{\WR}{\textsf{\textup{WR}}}

\newcommand{\CO}{\textsf{\textup{CO}}}
\newcommand{\CM}{\textsf{\textup{CM}}}

\providecommand{\wr}{}
\renewcommand{\wr}{\textsf{\textup{wr}}}

\newcommand{\rel}[1]{\xrightarrow{#1}}

\newcommand{\po}{{\sf po}}

\makeatletter
\newsavebox{\measure@tikzpicture}
\NewEnviron{scaletikzpicturetowidth}[1]{%
  \def\tikz@width{#1}%
  \begin{lrbox}{\measure@tikzpicture}%
  \BODY
  \end{lrbox}%
  \pgfmathparse{#1/\wd\measure@tikzpicture}%
  \BODY
}
\makeatother

\colorlet{sgray}{gray!80!black}
\colorlet{sgreen}{green!40!black}
\colorlet{sred}{red!70!black}
\colorlet{sbrown}{brown!80!black}
\colorlet{sblue}{blue!50!black}

\newcommand{\sorel}{\textsc{SO}}

\newcommand{\bpnotmysessionwriteco}{\textsc{NotMySessionWriteCO}}
\newcommand{\bpnotmysessionwritecm}{\textsc{NotMySessionWriteCM}}

\newcommand{\bpfracturedreadco}{\textsc{FracturedReadCO}}
\newcommand{\bpfracturedreadcm}{\textsc{FracturedReadCM}}

\newcommand{\rcone}{\textup{\texttt{RC1}\xspace}}
\newcommand{\rctwo}{\textup{\texttt{RC2}\xspace}}
\newcommand{\rcthree}{\textup{\texttt{RC3}\xspace}}
\newcommand{\rcfour}{\textup{\texttt{RC4}\xspace}}
\newcommand{\mav}{\textup{\texttt{MonoAtomicView}\xspace}}
\newcommand{\raaxiom}{\textup{\texttt{ReadAtomic}\xspace}}

\newcommand{\ReadX}{R_{x}}

\newcommand{\WriteX}{W_{x}}

\begin{document}

\title{Mechanizing Transactional Anomalous Patterns for Weak Isolation Levels}

\author
{Long Gu\inst{1}
\and
Si Liu\inst{2}
\thanks{Corresponding author.}
\and
Hengfeng Wei\inst{3}
}

\authorrunning{L. Gu et al.}

\institute{
Nanjing University
\and
The Chinese University of Hong Kong, Shenzhen
\and
Hunan University
}

\maketitle

\begin{abstract}

Recently proposed 
transactional anomalous patterns (TAPs)
 provide a semantic characterization of 
  weak 
 isolation levels  
 and
  serve as the foundation for TAP-based isolation checking.
Yet, these characterizations currently exist only as pen-and-paper proofs.
We present the first machine-checked formalization of TAPs and their associated weak isolation levels in Rocq. 
We further establish machine-checked proofs of the corresponding TAP-based characterization theorems.
The mechanization process revealed two subtle issues in the original presentation:
an underspecified transaction history model 
and
 an incomplete TAP characterization of the Read Atomicity isolation level that failed to capture violations of session guarantees. 
We address these issues by refining the definition of  transaction histories 
 and the corresponding TAP definitions, and re-establish all characterization theorems in Rocq.
These refinements 
have further been integrated into 
the weak  isolation checker Plume.

\end{abstract}

\section{Introduction}
\label{sec:intro}

Transactions are a fundamental abstraction in  database systems, 
 enabling applications to execute groups of operations atomically 
 despite concurrent access. 
	A key aspect of transactional correctness is \emph{isolation} 
	(the I in ACID~\cite{TIS:Book2002}), 
	which governs the effects of concurrent  transactions
  that may be observed by one another.
  To balance consistency and  performance~\cite{noc-noc,hat,NOC:OSDI2020,9139787}, 
   modern databases  support a spectrum of isolation levels, ranging from strong ones  such as \ser and \si
   to weaker ones  
   such as  \ci~\cite{hat}, \rc~\cite{critique}, \ra~\cite{RAMP:TODS2016}, and \tcc~\cite{Cure:ICDCS2016}. 
   These weaker levels are particularly important in
   \emph{highly available transactions} (HATs)~\cite{hat}, 
   as they enable always-on
   operation even in the presence of network partitions while still providing useful transactional semantics.

The importance of isolation guarantees has motivated extensive  research 
on black-box isolation checking~\cite{Complexity:OOPSLA2019,Cobra:OSDI2020,Elle:VLDB2020,PolySI:VLDB2023,Viper:EuroSys2023,IsoVista:VLDB24,Plume:OOPSLA2024,MiniTransactions:ICDE2025,AWDIT:PLDI2025,veristrong}, 
which determines whether an observed database execution of transactions
 satisfies a target isolation guarantee
  without  accessing  often unavailable or opaque database internals. 
 Among these checkers, 
Plume~\cite{Plume:OOPSLA2024},  recently developed by the authors and David Basin, 
focuses on HAT-compatible weak isolation levels.\footnote{Plume is part of a broader collaborative research effort on the reliability of database transactions from both the design~\cite{10.1145/3494517,noc-noc,VerIso:VLDB2025,EPplus,DBLP:conf/tase/LiangL21} and deployment~\cite{PolySI:VLDB2023,veristrong,MiniTransactions:ICDE2025,IsoVista:VLDB24} perspectives. The project is jointly led by the Information Security Group at ETH Zurich and its collaborators.}
One of its key contributions is the introduction of \emph{transactional anomalous patterns} (TAPs), 
 a collection of fine-grained anomalies that provide sound and complete characterizations of 
 HAT-compatible  isolation levels. 
 These characterization theorems
 constitute the  formal foundation of Plume: 
 its correctness relies on the equivalence between the absence of certain TAPs 
 and
  the satisfaction of  corresponding isolation guarantees.
 For example, in the original Plume characterization,
an execution history of transactions satisfies the \ra isolation level
if and only if  it excludes all instances of
12 TAPs (see, e.g., \cite[Fig.~3]{Plume:OOPSLA2024}).

However,
 like most formalizations of  isolation levels~\cite{Adya:PhDThesis1999,Framework:CONCUR2015,AnalysingSI:JACM2018,Seeing:PODC2017,Complexity:OOPSLA2019,Centralised:ECOOP2020,veristrong,PolySI:VLDB2023,Viper:EuroSys2023},
the equivalence theory underlying Plume  exists only as pen-and-paper proofs.
These proofs involve subtle reasoning about execution histories,
 transactional dependencies, isolation semantics, etc.,
 and often rely on implicit assumptions and omitted well-formedness conditions 
 that are not
  fully
  captured by the formal definitions. 
Hence, 
establishing a machine-checked formalization 
and
its 
equivalence results
 is valuable 
not only for increasing confidence in the proofs themselves,
 but also for validating the underlying definitions and their intended semantics.
 Otherwise, the claimed soundness 
(reporting no false positives) 
 and completeness
 (missing no isolation violations) 
  of TAP-based isolation checking in Plume
 cannot be guaranteed.

In this work, we present the first machine-checked formalization of TAPs
and their associated  weak isolation levels in Rocq~\cite{rocq}.
Specifically,
we mechanize the TAP theory,
including transactional histories, dependency relations,
the 16 TAPs (after refinement; see Section~\ref{sec:mech-taps}),
and the formal semantics of four HAT-compatible isolation levels, 
i.e.,
 \ci, \rc, \ra, and \tcc. 
We further establish machine-checked proofs of the soundness and completeness of the corresponding TAP-based characterization theorems.\footnote{Our Rocq development is 
	publicly available at \url{https://github.com/dracoooooo/Plume/tree/main/Mechanization}.}

Notably, the mechanization process revealed two subtle issues in the original presentation of Plume:

\begin{enumerate}
\item 
    The original transaction history model was underspecified, 
    relying on assumptions that were implicit in the pen-and-paper development but had to be made explicit for machine checking.
     Consequently, a direct mechanization was not possible without first refining the history model  (Section~\ref{subsec:mech-hist}).

\item 
    The original TAP characterization of \ra
     was semantically incomplete. 
     We identified a mismatch between two TAPs and the isolation semantics they were intended to characterize,  
     causing violations of session guarantees, specifically \emph{read my writes}~\cite{DBLP:journals/cacm/Terry13},
      to go undetected. 
     We address this issue by refining the corresponding TAPs
       and re-establishing the characterization theorems (Section~\ref{ss:verify}). 
      These refinements have subsequently been integrated into the Plume isolation checker.

\end{enumerate}

Overall, these findings highlight that mechanization is valuable not only for establishing high assurance in formal proofs, but also for exposing hidden assumptions, clarifying definitions, and improving the underlying theory.

\begin{figure}[t]
	\centering
	\includegraphics[width = \columnwidth]{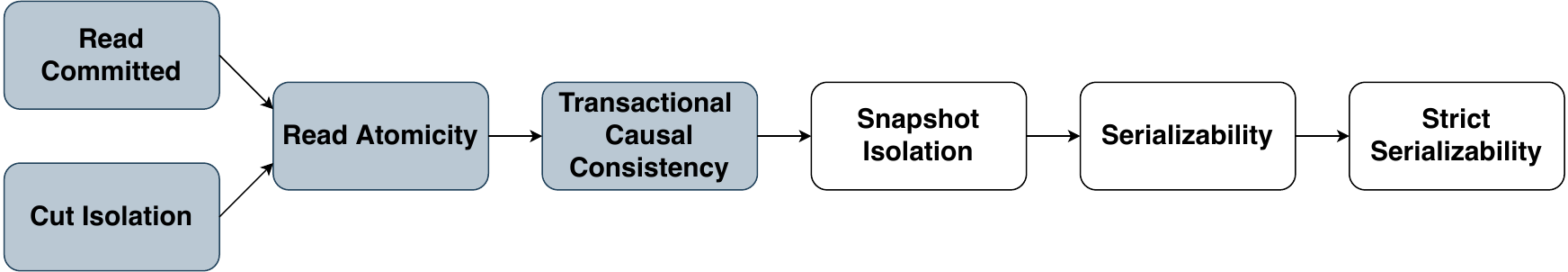}
	\captionsetup{skip=10pt}
	\caption{A hierarchy of  prevalent database isolation levels. The four HAT-compatible ones are highlighted.
		$I \to I'$: $I$ is strictly weaker than $I'$.
	}
	\label{fig:iso}
\end{figure}

\section{HAT-Compatible Weak Isolation Levels}

Modern database systems support a range of isolation levels (or guarantees),
 reflecting different trade-offs between consistency and performance~\cite{noc-noc,hat,NOC:OSDI2020,9139787}.
 In particular, 
 a class of weaker isolation guarantees
 plays a central role in 
supporting \emph{highly available transactions} (HATs)~\cite{hat},
since they enable   systems to remain available under network partitions while preserving meaningful consistency guarantees. 
 In the following,
we briefly introduce their  informal definitions.


\inlsec{\ci (CI)~\cite{hat}} 
This level
requires each transaction to observe
an unchanging snapshot of the database throughout its execution.
Hence,  reading the same data item  multiple times must always return the same value, ruling out the anomaly \emph{non-repeatable reads}~\cite{critique}.

\inlsec{\rc (RC)~\cite{critique}}
This level
guarantees that transactions observe only committed updates.
It prevents \emph{dirty reads},
but still permits anomalies such as non-repeatable reads.

\inlsec{\ra (RA)~\cite{RAMP:TODS2016}}
This level
ensures that updates produced by a transaction are observed atomically:
 either all of them are visible or none are.
This eliminates
\emph{fractured reads},
 where a transaction observes only a subset of another transaction's updates.
For instance,
Charlie
only sees one direction of a new (bidirectional) friendship between
Alice and Bob in a social network. 

Note that Plume adopts a slightly stronger variant of  RA~\cite{Complexity:OOPSLA2019} 
by additionally enforcing the \emph{read my writes} session guarantee~\cite{DBLP:journals/cacm/Terry13}. Intuitively, this guarantee ensures that once a client successfully commits a transaction that writes some data, every subsequent transaction in the same client session observes those writes (or more recent values).

\inlsec{\tcc (TCC)~\cite{Cure:ICDCS2016}}
This level 
extends \ra with \emph{causal ordering} and \emph{convergence} guarantees.
Causally related transactions must be observed in the same order by all sessions,\footnote{
A client session represents a sequence of interactions issued by a client to the database over a period of time,
e.g., from login to logout.} 
 preventing causality violations
 (e.g., Alice seeing Bob's
 response to Charlie's message without seeing the message itself).
Moreover, all replicas are required to eventually converge to the same state despite concurrent updates.

\smallskip
The recently developed black-box checker Plume~\cite{Plume:OOPSLA2024} targets these HAT-compatible isolation levels.
 It is built on two key ideas,
 addressing correctness and efficiency of checking, respectively.
  First, it introduces 
  \emph{transactional anomalous patterns} (TAPs),
  which provide sound and complete semantic characterizations of these isolation levels.
   Second, it combines vectors and tree clocks to efficiently 
   detect TAP instances in the  transactional dependency graph~\cite{Complexity:OOPSLA2019}  derived from an execution history,
    thereby accelerating isolation checking. 

In this work, we focus on the former aspect,
which underpins
 the 
correctness
 of TAP-based isolation checking.



\section{Mechanizing Transaction Histories and Dependencies}
\label{sec:mech-basic}

In this section, we mechanize in Rocq the notions of transactions, histories, and transactional dependencies
that underpin the mechanization of isolation levels (Section~\ref{sec:mech-iso}) and TAPs (Section~\ref{sec:mech-taps}).
\subsection{Notation}
\label{ss:notation}

A binary relation $R$ over a set $A$ is a subset of $A \times A$. 
For $a, b \in A$, we write $(a, b) \in R$ and $a \rel{R} b$ interchangeably.
Let $I_{A} \triangleq \set{(a, a) \mid a \in A}$
be the identity relation on $A$.
We use $R^{+}$ to denote the transitive closure of $R$.
A relation $R \subseteq A \times A$ is {\emph{acyclic}} if $R^{+} \cap I_{A} = \emptyset$.
A strict partial order is an irreflexive and transitive relation.
A strict total order is a relation that is a strict partial order and total.

In logical formulas,
we often write `$\_$' for components that are irrelevant
and implicitly existentially quantified,
and $\exists!$ to mean ``unique existence.''

Our mechanization is developed in Rocq~\cite{rocq},
an interactive theorem prover based on dependent type theory.
In the excerpts that follow,
computational objects such as operations and transactions
are assigned type \texttt{Type},
whereas specifications and lemmas are stated as propositions of type \texttt{Prop}.
An \texttt{Inductive} declaration introduces a datatype by listing its constructors
(e.g., \texttt{Read} and \texttt{Write} for operations).
A \texttt{Record} bundles named components into a single structured object;
besides data fields, a record may also carry \texttt{Prop}-typed fields
that encode well-formedness constraints or axioms
(e.g., \texttt{wr\_unique} and \texttt{so\_wr\_acyclic} in \texttt{History}).
Named concepts are introduced with \texttt{Definition}.

Sets are represented as \texttt{Ensemble}s,
i.e., predicates over a carrier type:
membership \texttt{E a} means that \texttt{a} belongs to the set \texttt{E}.
A binary relation over \texttt{A} has type \texttt{relation A},
equivalently \texttt{A -> A -> Prop};
we write \texttt{R a b} when $(a,b) \in R$.
Relations are often given by anonymous functions \texttt{fun a b => \dots}.
The standard library provides reusable order-theoretic notions
such as \texttt{strict\_order} and the transitive closure \texttt{clos\_trans}.

Logical formulas use universal quantification (\texttt{forall}),
existential quantification (\texttt{exists}),
unique existence (\texttt{exists!}),
and the usual connectives for conjunction, disjunction,
implication (\texttt{->}), and bi-implication (\texttt{<->}).
Record fields are accessed by projection:
for a history \texttt{H}, \texttt{T H} denotes its transaction set
and \texttt{WR H x} its write-read relation on key \texttt{x}.
Isolation levels and TAPs are defined as \texttt{Prop}s
that quantify over histories, commit orders, and transactions.

\subsection{Mechanizing Operations and Transactions}
\label{ss:ops-txns}

We consider a key-value data store managing a set of keys
$\Key = \set{x, y, z, \dots}$ associated with values from a set $\Val$.
We use $\readevent(x, v)$ to denote a read operation
that reads $v \in \Val$ from $x \in \Key$
and $\writeevent(x, v)$ to denote a write operation
that writes $v \in \Val$ to $x \in \Key$.
Each operation has a unique identifier.
A {\it transaction} is a pair $(O,\po)$,
where $O$ is a set of read and write operations,
each with its own unique identifier,
and $\po$ is a strict total order over the operations in $O$.
Figure~\ref{fig:rocq-ops-txns} shows the Rocq excerpt
for operations and transactions.

\begin{figure}[h]
	\input{code/op}
	\input{code/txn}
	\caption{Rocq excerpt: operations and transactions.}
	\label{fig:rocq-ops-txns}
\end{figure}

We use $t, t_{1}, t_{2}, \ldots$ to range over transactions,
and  $\Read(t)$, $\ReadX(t)$,
$\Write(t)$, and $\WriteX(t)$ to denote
the sets of read operations, read operations on $x$,
write operations, and write operations on $x$
in a transaction $t$, respectively.
The extension to sets of transactions is straightforward.
For example, $\Read(T) = \bigcup_{t \in T} \Read(t)$
denotes the set of read operations in a set $T$ of transactions.
We also use $\ReadTx_{x}$ and $\WriteTx_{x}$ to denote
the sets of transactions that read and write to $x$, respectively.
In Rocq, the predicate
{\texttt{Wx\;t\;x\;w}} (resp. {\texttt{Rx\;t\;x\;r}})
is true if and only if $w$ (resp. $r$)
is a write (resp. read) operation on key $x$ from transaction $t$,
namely, $w \in \WriteX(t)$ (resp. $r \in \ReadX(t)$).
\subsection{Mechanizing Histories}
\label{subsec:mech-hist}

A history records the  interactions between clients and the database,
including the set $T$ of transactions issued by clients,
the {\it session order} in which transactions are issued by each client,
and the {\it write-read relation} between operations.

\inlsec{Plume's Definition of History}
The original definition of history used by Plume
builds its write-read relation at the {\it transaction} level,
focusing on the {\it external} reads of transactions.
Specifically, for a transaction $t$,
we write $t \vdash \writeevent(x, v)$
if $t$ writes to $x$ and the last value written is $v$,
and $t \vdash \readevent(x, v)$
if $t$ has a read of $x$ returning $v$
that precedes every write to $x$ in $t$
(and hence also if $t$ performs no write to $x$),
and $v$ is the value returned by the {\it first} such read.
Then, the transaction-level write-read relation $\WR$ is defined as follows:
\begin{itemize}
  \item $\forall t, s \in T.\; \forall x \in \Key.\;
 		t \rel{\WR(x)} s \implies \exists v \in \Val.\;
 		t \neq s \land t \vdash \writeevent(x, v) \land s \vdash \readevent(x, v)$; and
 	\item $\forall s \in T.\; \forall x \in \Key.\; s \vdash \readevent(x, \_)
 		\iff \exists!\; t \in T.\; t \rel{\WR(x)} s$.
\end{itemize}

The first condition types each $\WR(x)$ edge,
and the second ensures that every external
read, and only an external read,
has exactly one source transaction.
Note that without the unique value assumption
under which for each key
every write to the key assigns a unique value,
the reverse implication in the first condition may not hold.

For convenience, we abuse notation and use
$\WR$ to also denote the relation
$\bigcup_{x \in \Key} \WR(x) \subseteq T \times T$.

\begin{definition}[History~\cite{Plume:OOPSLA2024}]
  \label{def:plume-history}
 	A history is a tuple $\h = (T, \SO, \WR)$,
 	where $T$ is a set of committed transactions,
 	$\sorel \subseteq T \times T$ is the {session order},
 	and $\WR : \Key \to 2^{T \times T}$ is the write-read relation,
 	such that $(\SO \cup \WR)$ is acyclic.
 \end{definition}
%

\inlsec{Refining the Original  History Definition}
During mechanization,
 we identified a limitation in Plume's definition of histories.
  Specifically, its transaction-level write-read relation $\WR$ is incomplete with respect to the operation-level write-read relation $\wr$,
 which requires every read operation $r$ to be associated with a unique write operation $w$
  on the same key and value.
  Consequently, $\WR$ cannot, in general, be derived from $\wr$ in the presence of internal reads.

  This limitation hinders the mechanization of the proofs of the implications between isolation levels, as well as
  the soundness and completeness proofs of the TAP-based characterization theorems.
Therefore,
in the new definition of history below,
we extend $\WR$ to be equivalent to $\wr$.

\begin{definition}[History (this work)]
	\label{def:history}
	A history is a tuple $\h = (T, \SO, \WR, \wr)$,
  where $T$ is the set of committed transactions,
  $\SO \subseteq T \times T$ is the session order,
  $\WR : \Key \to 2^{T \times T}$
  is the transaction-level write-read relation,
  and $\wr: \Key \to 2^{\Write(T) \times \Read(T)}$
  is the operation-level write-read relation such that
  \begin{itemize}
    \item If $w \rel{\wr(x)} r$,
      then $w$ writes to the same key and value as $r$ reads.
      \begin{align*}
        &\forall x \in \Key.\;
        \forall r \in \Read(T).\;
        \forall w \in \Write(T).\; \\
        &\quad (w \rel{\wr(x)} r \implies \exists v \in \Val.\;
          w = \writeevent(x, v) \land r = \readevent(x, v)).
      \end{align*}
    \item For each key $x$, every read operation $r$ on $x$
      is associated with a unique write operation $w$
      on $x$ such that $w \rel{\wr(x)} r$.
      \[
        \forall x \in \Key.\;
        \forall r \in \ReadX(T).\;
        \exists! w \in \WriteX(T).\;
        w \rel{\wr(x)} r.
      \]
    \item $\WR$ and $\wr$ can be derived from each other as follows:
      \begin{align*}
        \forall x \in \Key.\; \forall t, s \in T.\;
        (t \rel{\WR(x)} s \iff t \neq s \land
        \exists w \in \WriteX(t), r \in \ReadX(s).\; w \rel{\wr(x)} r).
      \end{align*}
    \item $(\SO \cup \WR)$ is acyclic.
  \end{itemize}
\end{definition}

We assume a distinguished initial transaction
that writes the initial value of every key to the database.
Figure~\ref{fig:rocq-history} shows
the corresponding Rocq excerpt.
The four conditions in Definition~\ref{def:history}
correspond to the definitions of \texttt{wr}
(as a global relation on operations
constrained by the \texttt{History} record;
see lines 1--3),
\texttt{wr\_unique}
(lines 10--12),
\texttt{wr\_iff\_WR}
(lines 14--17)
and \texttt{so\_wr\_acyclic}
(lines 19--22), respectively.

\begin{figure}[htb]
	\input{code/history}
	\caption{Rocq excerpt: the refined definition of history.}
	\label{fig:rocq-history}
		\vspace{-2ex}
\end{figure}

\begin{figure}[t]
	\begin{lstlisting}[style=coqstyle]
		Definition SO_union_WR (H : History) : relation Transaction :=
		fun t1 t2 => SO H t1 t2 \/ exists x, WR H x t1 t2.

		Definition SO_union_WR_plus (H : History) : relation Transaction :=
		clos_trans Transaction (SO_union_WR H).

		Definition CO (H : History) : relation Transaction :=
		SO_union_WR_plus H.

		Definition commit_order
		(H : History) (CM : relation Transaction) : Prop :=
		strict_order CM /\
		(forall t1 t2,
		T H t1 -> T H t2 -> t1 <> t2 ->
		CM t1 t2 \/ CM t2 t1) /\
		(forall t1 t2, CO H t1 t2 -> CM t1 t2).
	\end{lstlisting}
	\caption{Rocq excerpt: the causal order and the commit order.}
	\label{fig:rocq-order}
\end{figure}



\subsection{Mechanizing Transactional Dependencies}
\label{ss:txn-deps}

A history satisfies an isolation level if
there exists a strict total order over its transactions,
called the {\it commit order}~\cite{Complexity:OOPSLA2019},
which extends the causal order,
i.e., the transitive closure of the union of
the write-read relation and the session order,
under which the axioms defining the isolation level hold.
Intuitively, the commit order dictates the order in which
transactions are committed to the ``global'' data store.

We define the {\it causal order} $\CO$ as
the transitive closure of $\SO$ and $\WR$,
i.e., $\CO \triangleq (\SO \cup \WR)^{+}$,
capturing the potential causality between transactions.
A commit order $\CM \subseteq T \times T$
is a strict total order that extends $\CO$.
Figure~\ref{fig:rocq-order}  presents the corresponding  definitions in Rocq.
\section{Mechanizing HAT-Compatible Isolation Levels}
\label{sec:mech-iso}

Building on the formalization of histories and ordering relations,
 this section presents the Rocq formalization of the four HAT-compatible isolation levels.


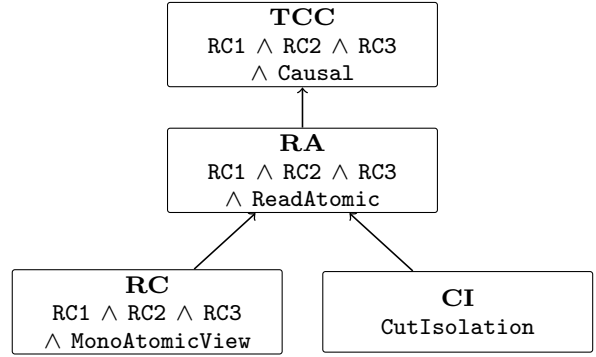
\begin{figure}[t]
	\centering
	\begin{tikzpicture}[
		    scale=1.3,
		transform shape,
		defn/.style={
			draw,
			rounded corners=1pt,
			align=center,
			inner sep=1.8pt,
			font=\scriptsize,
			text width=2.62cm,
			minimum height=0.82cm
		},
		arr/.style={->, semithick}
	]
	\node[defn] (tcc) at (0,0)
		{\textbf{TCC}\\
		\texttt{RC1} $\land$ \texttt{RC2} $\land$ \texttt{RC3}\\
		$\land$ \texttt{Causal}};
	\node[defn] (ra) at (0,-1.28)
		{\textbf{RA}\\
		\texttt{RC1} $\land$ \texttt{RC2} $\land$ \texttt{RC3}\\
		$\land$ \texttt{ReadAtomic}};
	\node[defn] (rc) at (-1.58,-2.72)
		{\textbf{RC}\\
		\texttt{RC1} $\land$ \texttt{RC2} $\land$ \texttt{RC3}\\
		$\land$ \texttt{MonoAtomicView}};
	\node[defn] (ci) at (1.58,-2.72)
		{\textbf{CI}\\
		\texttt{CutIsolation}};

	\draw[arr] (ra) -- (tcc);
	\draw[arr] (rc) -- (ra);
	\draw[arr] (ci) -- (ra);
	\end{tikzpicture}
	\caption{Definitions of the four HAT-compatible isolation levels.
An edge $A \rightarrow B$ indicates that $A$ is strictly weaker than $B$.}
	\label{fig:levels}
\end{figure}


As shown in Figure~\ref{fig:levels},
we mechanize seven axioms in total, namely \texttt{RC1}, \texttt{RC2}, \texttt{RC3}, \texttt{CutIsolation}, \texttt{MonoAtomicView}, \texttt{ReadAtomic}, and \texttt{Causal},
 together with the four HAT-compatible isolation levels \rc, \ci, \ra, and \tcc,
 each defined as a conjunction of these axioms.

In the following, we use \ra as a case study, since mechanizing it revealed another subtle issue in its TAP-based characterization
(Section~\ref{ss:verify}).
Intuitively, \ra guarantees that a transaction's updates become visible as a single atomic unit.
Hence, another transaction can either observe all of those updates or none of them, ruling out \emph{fractured reads}.

\begin{definition}[\ra~\cite{Plume:OOPSLA2024}]
	\label{def:ra}
	A history $\h = (T, \SO, \WR, \wr)$ satisfies \ra
	if and only if it satisfies \rc
  (specifically, \rcone, \rctwo, \rcthree, and \rcfour{} (i.e., \mav))
  and there exists a commit order $\CM \subseteq T \times T$ on $T$
	such that the following \raaxiom{} axiom holds.

	(\raaxiom{} axiom)
	\emph{If a transaction $t_{3}$ reads a key $x$ from $t_{1}$
		and there exists a transaction $t_{2} \neq t_{1}$
		that also writes to $x$ such that $t_{2} \rel{\SO \;\cup\; \WR} t_{3}$,
		then $t_{2} \rel{\CM} t_{1}$.}
	Formally, $\forall x \in \Key.\;
	\forall t_{1}, t_{2}  \in \WriteTx_{x}.\;
	\forall t_{3} \in \ReadTx_{x} \setminus \set{t_{1}, t_{2}}.\;
	(t_{1} \neq t_{2} \land t_{1} \rel{\WR(x)} t_{3}
	\land t_{2} \rel{\SO \;\cup\; \WR} t_{3} \implies t_{2} \rel{\CM} t_{1})$.
\end{definition}

\begin{figure}[tbhp]
  \input{code/ra}
  \caption{Rocq excerpt: the \raaxiom{} axiom and \ra.}
  \label{fig:rocq-ra}
\end{figure}

Figure~\ref{fig:rocq-ra} shows the formalization
of \raaxiom{} axiom and the \ra{} isolation level in Rocq.
Specifically,
the predicate \texttt{WTx (T H) x t1} (line 4)
(resp., \texttt{RTx (T H) x t3} (line 6)) is true if and only if
transaction $t_1$ (resp., $t_3$) writes (resp., reads) key $x$.
The edge \texttt{WR H x t1 t3} means that
$t_3$ reads key $x$ from $t_1$ (line 7).
The edge \texttt{SO\_union\_WR H t2 t3} means that
$t_2$ is visible to $t_3$ through one $\SO\cup\WR$ step (line 8).
The conclusion \texttt{CM t2 t1} requires that
$t_2$ must be ordered before $t_1$ in the commit order (line 9).

We omit the definitions of \rcone, \rctwo, and \rcthree, and note that \rcfour{} (i.e., \mav) is implied by \raaxiom.
The complete mechanized definitions,
including those for other isolation levels,
 can be found in our repository.

\section{Mechanizing TAPs and Their Semantic Conformance}
\label{sec:mech-taps}

This section mechanizes the
transactional anomalous patterns (TAPs), focusing on those for \ra,
and verifies the equivalence between the absence of  corresponding TAPs and the formal definition of \ra given in Section~\ref{sec:mech-iso}.
Our initial attempt to establish this equivalence failed,
 revealing a subtle semantic issue in the original TAPs:
  they do not capture anomalies arising from violations of the \emph{read my writes}  session guarantee~\cite{DBLP:journals/cacm/Terry13}.
   This observation led us to refine the TAP-based characterization and subsequently complete the machine-checked proof of equivalence.

\subsection{Mechanizing TAPs}
\label{ss:mech-taps}


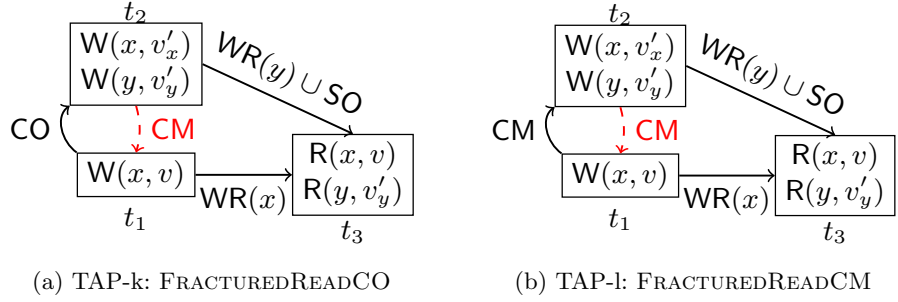
\begin{figure}[t]
	\centering
	\begin{subfigure}[b]{0.46\linewidth}
		\centering
		\resizebox{\linewidth}{!}{%
			\begin{tikzpicture}[
				op/.style={draw, inner xsep=3pt, inner ysep=2.2pt, align=center},
				edge/.style={->, line width=0.6pt},
				inferred/.style={edge, red, dashed},
				font=\small
				]
				\node[op] (t2w) at (0,1.2) {$\writeevent(x,v_{x}')$\\$\writeevent(y,v_{y}')$};
				\node[op] (t1w) at (0,0) {$\writeevent(x,v)$};
				\node[op] (t3r) at (2.35,0) {$\readevent(x,v)$\\$\readevent(y,v_{y}')$};

				\node at (0,1.76) {$t_2$};
				\node at (0,-0.48) {$t_1$};
				\node at (2.35,-0.62) {$t_3$};

				\draw[edge] (t1w.east) -- node[below,sloped] {$\WR(x)$} (t3r.west);
				\draw[edge] (t2w.east) -- node[above,sloped] {$\WR(y)\cup\SO$} (t3r.north);
				\draw[edge] (t1w.north west) to[out=145,in=215] node[left] {$\CO$} (t2w.south west);
				\draw[inferred] (t2w.south) to[out=-75,in=75] node[right,red] {$\CM$} (t1w.north);
			\end{tikzpicture}
		}
		\caption{TAP-k: \bpfracturedreadco}
	\end{subfigure}
	\qquad
	\begin{subfigure}[b]{0.46\linewidth}
		\centering
		\resizebox{\linewidth}{!}{%
			\begin{tikzpicture}[
				op/.style={draw, inner xsep=3pt, inner ysep=2.2pt, align=center},
				edge/.style={->, line width=0.6pt},
				inferred/.style={edge, red, dashed},
				font=\small
				]
				\node[op] (t2w) at (0,1.2) {$\writeevent(x,v_{x}')$\\$\writeevent(y,v_{y}')$};
				\node[op] (t1w) at (0,0) {$\writeevent(x,v)$};
				\node[op] (t3r) at (2.35,0) {$\readevent(x,v)$\\$\readevent(y,v_{y}')$};

				\node at (0,1.76) {$t_2$};
				\node at (0,-0.48) {$t_1$};
				\node at (2.35,-0.62) {$t_3$};

				\draw[edge] (t1w.east) -- node[below,sloped] {$\WR(x)$} (t3r.west);
				\draw[edge] (t2w.east) -- node[above,sloped] {$\WR(y)\cup\SO$} (t3r.north);
				\draw[edge] (t1w.north west) to[out=145,in=215] node[left] {$\CM$} (t2w.south west);
				\draw[inferred] (t2w.south) to[out=-75,in=75] node[right,red] {$\CM$} (t1w.north);
			\end{tikzpicture}
		}
		\caption{TAP-l: \bpfracturedreadcm}
	\end{subfigure}

	\caption{The two original TAPs for \emph{fractured read}~\cite{Plume:OOPSLA2024}.
	  Solid black arrows are edges present in the TAPs.
		Red dashed arrows are commit-order edges entailed by the \raaxiom{} axiom;
    together with the solid commit-order edge,
		they expose the contradiction.}
	\label{fig:tap-kl-original}
\end{figure}

\begin{figure}[t]
	\centering
	\begin{bluetapbox}{TAP-k: \bpfracturedreadco}
		\[
		\begin{aligned}
			&\exists x, y \neq x \in \Key.\;
			\exists t_1, t_2 \neq t_1 \in \WriteTx_x.\;
			\exists t_3 \in (\ReadTx_x \cap \ReadTx_y)
			\setminus \set{t_1,t_2}.\\
			&\exists w_x \in W_x(t_1).\;
			\exists w_y \in W_y(t_2).\;
			\exists r_x \in R_x(t_3).\;
			\exists r_y \in R_y(t_3).\\
			&\left(
			w_x \rel{\wr(x)} r_x
			\land
			(w_y \rel{\wr(y)} r_y \lor t_{2} \rel{\SO} t_{3})
			\land
			t_1 \rel{\CO} t_2
			\right).
		\end{aligned}
		\]
	\end{bluetapbox}

	\begin{bluetapbox}{TAP-l: \bpfracturedreadcm}
		\[
		\begin{aligned}
			&\exists x, y \neq x \in \Key.\;
			\exists t_1, t_2 \neq t_1 \in \WriteTx_x.\;
			\exists t_3 \in (\ReadTx_x \cap \ReadTx_y)
			\setminus \set{t_1,t_2}.\\
			&\exists w_x \in W_x(t_1).\;
			\exists w_y \in W_y(t_2).\;
			\exists r_x \in R_x(t_3).\;
			\exists r_y \in R_y(t_3).\\
			&\left(
			w_x \rel{\wr(x)} r_x
			\land
			(w_y \rel{\wr(y)} r_y \lor t_{2} \rel{\SO} t_{3})
			\land
			t_1 \rel{\CM} t_2
			\right).
		\end{aligned}
		\]
	\end{bluetapbox}
	\caption{Formal definitions of the two original TAPs in Figure~\ref{fig:tap-kl-original}.}
	\label{fig:tap-kl-original-formal}
\end{figure}

Plume devises 14 fine-grained TAPs
to characterize the four HAT-compatible isolation levels.
\ra
is defined using 12 of these TAPs:
nine are shared with \rc, one with \ci, and the remaining two capture fractured reads, the distinguishing anomaly prohibited by \ra.
Figure~\ref{fig:tap-kl-original} shows these
two TAPs, and  their formalization is given in Figure~\ref{fig:tap-kl-original-formal}.\footnote{
We note that the textual descriptions and formal definitions of
 TAP-k and TAP-l in~\cite[Table 1]{Plume:OOPSLA2024}
	are inconsistent with their visualizations
	in~\cite[Figure 3]{Plume:OOPSLA2024},
		missing the required $\SO$ edges.
 Figure~\ref{fig:tap-kl-original} shows our corrected versions.
}

Specifically,
in TAP-k (\bpfracturedreadco),
transaction $t_{3}$ reads $x$ from $t_{1}$
and $y \neq x$ from $t_{2}$,
while transaction $t_{2}$ also writes to $x$ but $t_{1} \rel{\CO} t_{2}$.
Similarly,
in TAP-l (\bpfracturedreadcm),
transaction $t_{3}$ reads $x$ from $t_{1}$ and  $y \neq x$ from $t_{2}$,
while transaction $t_{2}$ also writes to $x$ but $t_{1} \rel{\CM} t_{2}$.
In both cases,
$t_{2}$ should precede $t_{1}$ in $\CM$,
contradicting $t_{1} \rel{\CO} t_{2}$ in TAP-k
and $t_{1} \rel{\CM} t_{2}$ in TAP-l, respectively.
Note that, since $\CO \subseteq \CM$,
TAP-l is a general case of TAP-k.
We retain both named patterns
for diagnosis and comparison with Plume~\cite{Plume:OOPSLA2024},
which aims to provide a fine-grained characterization of anomalies.
The proof basis is minimal:
because TAP-k implies TAP-l, the absence of TAP-l suffices.
This also applies to TAP-o and TAP-p,
which we introduce in Section~\ref{ss:verify}.

\subsection{Initial Attempt to Prove the Equivalence}
\label{ss:verify}


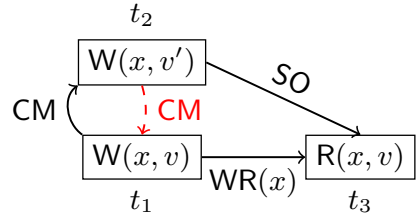
\begin{figure}[t]
	\centering
	\resizebox{0.46\linewidth}{!}{%
		\begin{tikzpicture}[
			op/.style={draw, inner xsep=3pt, inner ysep=2.2pt, align=center},
			edge/.style={->, line width=0.6pt},
			inferred/.style={edge, red, dashed},
			font=\small
		]
		\node[op] (t2w) at (0,1.02) {$\writeevent(x,v')$};
		\node[op] (t1w) at (0,0) {$\writeevent(x,v)$};
		\node[op] (t3r) at (2.35,0) {$\readevent(x,v)$};

		\node at (0,1.53) {$t_2$};
		\node at (0,-0.48) {$t_1$};
		\node at (2.35,-0.48) {$t_3$};

		\draw[edge] (t1w.east) -- node[below] {$\WR(x)$} (t3r.west);
		\draw[edge] (t2w.east) -- node[above,sloped] {$\SO$} (t3r.north);
		\draw[edge] (t1w.north west) to[out=145,in=215] node[left] {$\CM$} (t2w.south west);
		\draw[inferred] (t2w.south) to[out=-75,in=75] node[right,red] {$\CM$} (t1w.north);
	\end{tikzpicture}
	}
	\caption{A history violating \ra{}, but not captured
		by Plume's original TAP-k or TAP-l.}
	\label{fig:bug}
\end{figure}


\begin{figure}[t]
	\centering
	\begin{subfigure}[b]{0.46\linewidth}
		\centering
		\resizebox{\linewidth}{!}{%
			\begin{tikzpicture}[
				op/.style={draw, inner xsep=3pt, inner ysep=2.2pt, align=center},
				edge/.style={->, line width=0.6pt},
				inferred/.style={edge, red, dashed},
				font=\small
			]
				\node[op] (t2w) at (0,1.2) {$\writeevent(x,v_{x}')$\\$\writeevent(y,v_{y}')$};
				\node[op] (t1w) at (0,0) {$\writeevent(x,v)$};
				\node[op] (t3r) at (2.35,0) {$\readevent(x,v)$\\$\readevent(y,v_{y}')$};

				\node at (0,1.76) {$t_2$};
				\node at (0,-0.48) {$t_1$};
				\node at (2.35,-0.62) {$t_3$};

				\draw[edge] (t1w.east) -- node[below,sloped] {$\WR(x)$} (t3r.west);
				\draw[edge] (t2w.east) -- node[above,sloped] {$\WR(y)$} (t3r.north);
				\draw[edge] (t1w.north west) to[out=145,in=215] node[left] {$\CO$} (t2w.south west);
				\draw[inferred] (t2w.south) to[out=-75,in=75] node[right,red] {$\CM$} (t1w.north);
			\end{tikzpicture}
		}
		\caption{New TAP-k: \bpfracturedreadco}
	\end{subfigure}
	\qquad
	\begin{subfigure}[b]{0.46\linewidth}
		\centering
		\resizebox{\linewidth}{!}{%
			\begin{tikzpicture}[
				op/.style={draw, inner xsep=3pt, inner ysep=2.2pt, align=center},
				edge/.style={->, line width=0.6pt},
				inferred/.style={edge, red, dashed},
				font=\small
			]
				\node[op] (t2w) at (0,1.2) {$\writeevent(x,v_{x}')$\\$\writeevent(y,v_{y}')$};
				\node[op] (t1w) at (0,0) {$\writeevent(x,v)$};
				\node[op] (t3r) at (2.35,0) {$\readevent(x,v)$\\$\readevent(y,v_{y}')$};

				\node at (0,1.76) {$t_2$};
				\node at (0,-0.48) {$t_1$};
				\node at (2.35,-0.62) {$t_3$};

				\draw[edge] (t1w.east) -- node[below,sloped] {$\WR(x)$} (t3r.west);
				\draw[edge] (t2w.east) -- node[above,sloped] {$\WR(y)$} (t3r.north);
				\draw[edge] (t1w.north west) to[out=145,in=215] node[left] {$\CM$} (t2w.south west);
				\draw[inferred] (t2w.south) to[out=-75,in=75] node[right,red] {$\CM$} (t1w.north);
			\end{tikzpicture}
		}
		\caption{New TAP-l:  \bpfracturedreadcm}
	\end{subfigure}

	\begin{subfigure}[b]{0.46\linewidth}
		\centering
		\resizebox{\linewidth}{!}{%
			\begin{tikzpicture}[
			op/.style={draw, inner xsep=3pt, inner ysep=2.2pt, align=center},
			edge/.style={->, line width=0.6pt},
			inferred/.style={edge, red, dashed},
			font=\small
		]
			\node[op] (t2w) at (0,1.02) {$\writeevent(x,v')$};
			\node[op] (t1w) at (0,0) {$\writeevent(x,v)$};
			\node[op] (t3r) at (2.35,0) {$\readevent(x,v)$};

			\node at (0,1.53) {$t_2$};
			\node at (0,-0.48) {$t_1$};
			\node at (2.35,-0.48) {$t_3$};

			\draw[edge] (t1w.east) -- node[below] {$\WR(x)$} (t3r.west);
			\draw[edge] (t2w.east) -- node[above,sloped] {$\SO$} (t3r.north);
			\draw[edge] (t1w.north west) to[out=145,in=215] node[left] {$\CO$} (t2w.south west);
			\draw[inferred] (t2w.south) to[out=-75,in=75] node[right,red] {$\CM$} (t1w.north);
		\end{tikzpicture}
		}
		\caption{TAP-o: \bpnotmysessionwriteco}
	\end{subfigure}
	\qquad
	\begin{subfigure}[b]{0.46\linewidth}
		\centering
		\resizebox{\linewidth}{!}{%
		\begin{tikzpicture}[
			op/.style={draw, inner xsep=3pt, inner ysep=2.2pt, align=center},
			edge/.style={->, line width=0.6pt},
			inferred/.style={edge, red, dashed},
			font=\small
		]
			\node[op] (t2w) at (0,1.02) {$\writeevent(x,v')$};
			\node[op] (t1w) at (0,0) {$\writeevent(x,v)$};
			\node[op] (t3r) at (2.35,0) {$\readevent(x,v)$};

			\node at (0,1.53) {$t_2$};
			\node at (0,-0.48) {$t_1$};
			\node at (2.35,-0.48) {$t_3$};

			\draw[edge] (t1w.east) -- node[below] {$\WR(x)$} (t3r.west);
			\draw[edge] (t2w.east) -- node[above,sloped] {$\SO$} (t3r.north);
			\draw[edge] (t1w.north west) to[out=145,in=215] node[left] {$\CM$} (t2w.south west);
			\draw[inferred] (t2w.south) to[out=-75,in=75] node[right,red] {$\CM$} (t1w.north);
		\end{tikzpicture}
		}
		\caption{TAP-p: \bpnotmysessionwritecm}
	\end{subfigure}

	\caption{Refined TAPs for \emph{fractured reads}
	  and  for violations of  the session guarantee \emph{read my writes}.}
	\label{fig:tap-kl-new}
\end{figure}
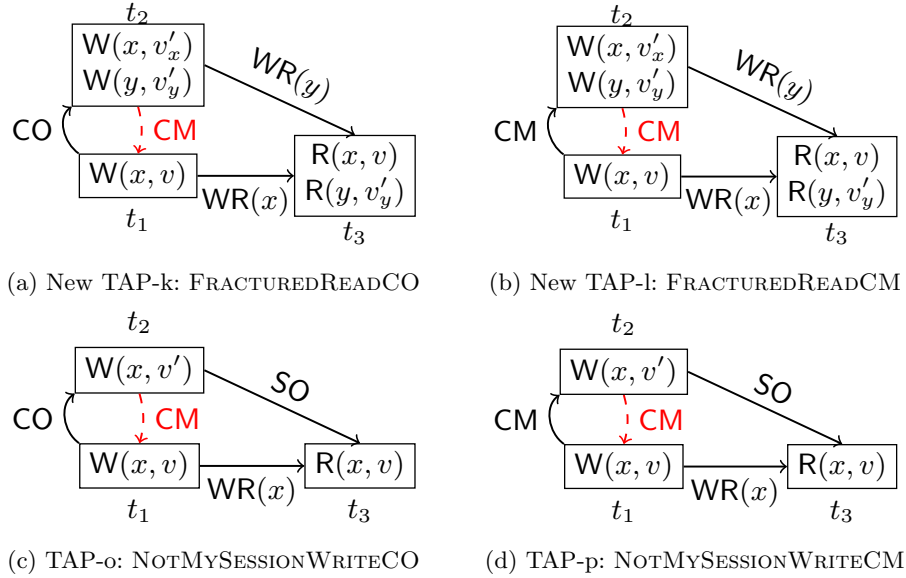

\begin{figure}[t]
	\centering

	\begin{greentapbox}{New TAP-k: \bpfracturedreadco}
		\[
		\begin{aligned}
			&\exists x, y \neq x \in \Key.\;
			\exists t_1, t_2 \neq t_1 \in \WriteTx_x.\;
			\exists t_3 \in (\ReadTx_x \cap \ReadTx_y)
			\setminus \set{t_1,t_2}.\\
			&\exists w_x \in W_x(t_1).\;
			\exists w_y \in W_y(t_2).\;
			\exists r_x \in R_x(t_3).\;
			\exists r_y \in R_y(t_3).\\
			&\left(
			w_x \rel{\wr(x)} r_x
			\land
			w_y \rel{\wr(y)} r_y
			\land
			t_1 \rel{\CO} t_2
			\right).
		\end{aligned}
		\]
	\end{greentapbox}

	\begin{greentapbox}{New TAP-l: \bpfracturedreadcm}
		\[
		\begin{aligned}
			&\exists x, y \neq x \in \Key.\;
			\exists t_1, t_2 \neq t_1 \in \WriteTx_x.\;
			\exists t_3 \in (\ReadTx_x \cap \ReadTx_y)
			\setminus \set{t_1,t_2}.\\
			&\exists w_x \in W_x(t_1).\;
			\exists w_y \in W_y(t_2).\;
			\exists r_x \in R_x(t_3).\;
			\exists r_y \in R_y(t_3).\\
			&\left(
			w_x \rel{\wr(x)} r_x
			\land
			w_y \rel{\wr(y)} r_y
			\land
			t_1 \rel{\CM} t_2
			\right).
		\end{aligned}
		\]
	\end{greentapbox}

	\begin{greentapbox}{TAP-o: {\bpnotmysessionwriteco}}
		\[
		\begin{aligned}
			&\exists x \in \Key.\;
			\exists t_1, t_2 \neq t_1 \in \WriteTx_x.\;
			\exists t_3 \in \ReadTx_x \setminus \set{t_1,t_2}.\\
			&\exists w_x \in W_x(t_1).\;
			\exists r_x \in R_x(t_3).\; \\
			&\left(
			w_x \rel{\wr(x)} r_x
			\land
			t_{2} \rel{\SO} t_{3}
			\land
			t_1 \rel{\CO} t_2
			\right).
		\end{aligned}
		\]
	\end{greentapbox}

	\begin{greentapbox}{TAP-p: {\bpnotmysessionwritecm}}
		\[
		\begin{aligned}
			&\exists x \in \Key.\;
			\exists t_1, t_2 \neq t_1 \in \WriteTx_x.\;
			\exists t_3 \in \ReadTx_x \setminus \set{t_1,t_2}.\\
			&\exists w_x \in W_x(t_1).\;
			\exists r_x \in R_x(t_3).\; \\
			&\left(
			w_x \rel{\wr(x)} r_x
			\land
			t_{2} \rel{\SO} t_{3}
			\land
			t_1 \rel{\CM} t_2
			\right).
		\end{aligned}
		\]
	\end{greentapbox}
	\caption{Formal definitions of the refined TAPs in Figure~\ref{fig:tap-kl-new}.}
	\label{fig:tap-kl-new-formal}
\end{figure}

\begin{figure}[t]
	\centering
		\begin{lstlisting}[style=coqstyle]
Definition TAP_l
    (H : History) (CM : relation Transaction) : Prop :=
  exists x t1 t2 t3,
    WTx (T H) x t1 /\ WTx (T H) x t2 /\ t1 <> t2 /\
    RTx (T H) x t3 /\ t3 <> t1 /\ t3 <> t2 /\
    (exists y, x <> y /\
      RTx (T H) y t3 /\ WR H x t1 t3 /\ CM t1 t2 /\ WR H y t2 t3).
		\end{lstlisting}

	\caption{Rocq excerpt:  the revised TAP-l  (\bpfracturedreadcm).}
	\label{fig:tap-l}
\end{figure}

\begin{figure}[t]
	\centering	
		\begin{lstlisting}[style=coqstyle]
Definition TAP_p 
		(H : History) (CM : relation Transaction) : Prop :=
	exists x t1 t2 t3,
		WTx (T H) x t1 /\ WTx (T H) x t2 /\ t1 <> t2 /\
		RTx (T H) x t3 /\ t3 <> t1 /\ t3 <> t2 /\
		(WR H x t1 t3 /\ CM t1 t2 /\ SO H t2 t3).
		\end{lstlisting}
	\caption{Rocq excerpt:  the new TAP-p  (\bpnotmysessionwritecm).}
	\label{fig:tap-p}
\end{figure}

Unfortunately, Plume's original TAP characterization of \ra is incomplete,
and
our initial attempt to prove the equivalence  failed.
Figure~\ref{fig:bug}
shows a history,
originally reported by Barros et al.~\cite{Isolde:arXiv2026},
 that violates \ra{}, specifically the \raaxiom{} axiom,
but is not captured by the corresponding TAPs, namely TAP-k and TAP-l.

In this history,
transaction $t_{1}$ is ordered before $t_{2}$ in $\CM$,
while $t_{2}$ is ordered before $t_{3}$ in $\SO$.
Transaction $t_{3}$ reads key $x$ from $t_{1}$,
while failing to observe the later write to $x$ performed by $t_{2}$.
The \raaxiom{} axiom requires $t_{2} \rel{\CM} t_{1}$,
contradicting $t_{1} \rel{\CM} t_{2}$.
Hence, this history violates \ra.
However, neither TAP-k nor TAP-l matches this history,
because
both patterns require at least
two read operations
 in $t_{3}$.

 The underlying issue is that the original TAP-k and TAP-l conflate two distinct kinds of anomalies, namely
\emph{fractured reads}
 and
violations of the session guarantee \emph{read my writes},
in a single pattern.
 To address this issue,
 we split each of the original patterns into two.
 The revised TAP-k (resp. TAP-l) captures only fractured reads,
 while the new TAP-o (\bpnotmysessionwriteco) and TAP-p (\bpnotmysessionwritecm)
 capture violations of
  \emph{read my writes}.
Figure~\ref{fig:tap-kl-new}
shows the refined TAPs,
and their formalization is given in Figure~\ref{fig:tap-kl-new-formal}.
Compared with the original definitions,
the revised TAP-k and TAP-l allow the case where
$t_{2} \rel{\WR} t_{3}$,
but exclude the case where
$t_{2} \rel{\SO} t_{3}$.
The latter is instead captured by TAP-o and TAP-p.
Since $\CO \subseteq \CM$,
TAP-p is a general case of TAP-o.

With these refinements,
the \raaxiom{} axiom is equivalent to the absence of
TAP-j (for  anomalies of \emph{non-repeatable reads}),
TAP-l, and TAP-p.
Figure~\ref{fig:tap-l} and Figure~\ref{fig:tap-p}
present the Rocq definitions of revised TAP-l and new TAP-p, respectively.

\subsection{Proving the Equivalence}
\label{subsec:verifying-equivalence}

Table~\ref{table:proof} summarizes the definitions of the four HAT-compatible isolation levels in terms of prohibited TAPs (16 in total after our refinement),
 together with the key lemmas used to prove their equivalence in Rocq.
It also highlights the modular structure of our proof development
 by showing how the characterization of each isolation level is built from a small set of reusable lemmas.
For example,
 the proof of \rc decomposes into lemmas establishing the equivalence between
	RC-(1--3)  and
	the absence of TAP-a--TAP-g,
	showing that TAP-h implies TAP-i,
	and finally connecting \texttt{MonoAtomicView} with the absence of TAP-i.
 Likewise, the proof of \ra builds on lemmas showing that TAP-k and TAP-o respectively imply TAP-l and TAP-p,
 before establishing the correspondence between
  \texttt{ReadAtomic} and
  the remaining TAPs
  and  semantic axioms.

 In total, our development comprises 1057 LOC in Rocq,
  736 LOC for proofs, 13 key lemmas, and 4  theorems.

\begin{table}[h]
	\centering
	\caption{Defining the four HAT-compatible isolation levels in terms of prohibited TAPs, together with the key lemmas used to prove their equivalence.}
	\label{table:proof}
	\small
	\setlength{\tabcolsep}{6pt}
	\begin{tabularx}{1\textwidth}{@{}c
			>{\raggedright\arraybackslash}X
			>{\raggedright\arraybackslash}p{0.50\textwidth}@{}}
		\toprule
		\textbf{Iso. Level} & \textbf{Prohibited TAPs} & \textbf{Key Lemmas} \\
		\midrule
		CI &
		$\text{TAP-j}$ &
		direct equivalence \\
		\midrule
		RC &
		$\text{TAP-a} \lor \cdots \lor \text{TAP-i}$ &
		RC-(1--3) $\Leftrightarrow \neg(\text{TAP-a} \lor \cdots \lor \text{TAP-g})$ \newline TAP-h $\Rightarrow$ TAP-i \newline \texttt{MonoAtomicView} $\Leftrightarrow \neg(\text{TAP-i})$ \\
		\midrule
		RA &
		$\text{TAP-a} \lor \cdots \lor \text{TAP-l} \lor \text{TAP-o} \lor \text{TAP-p}$ &
		TAP-k $\Rightarrow$ TAP-l \newline TAP-o $\Rightarrow$ TAP-p \newline \texttt{ReadAtomic} $\Leftrightarrow \neg(\text{TAP-j} \lor \text{TAP-l} \lor \text{TAP-p})$ \newline \texttt{ReadAtomic} $\Rightarrow$ CutIsolation \newline \texttt{ReadAtomic} $\Rightarrow$ \texttt{MonoAtomicView} \\
		\midrule
		TCC &
		$\text{TAP-a} \lor \cdots \lor \text{TAP-p}$ &
		\texttt{Causal} $\Rightarrow$ \texttt{ReadAtomic} \newline TAP-m $\Rightarrow$ TAP-n \newline \texttt{Causal} $\Leftrightarrow \neg(\text{TAP-n})$ \\
		\bottomrule
	\end{tabularx}
\end{table}




\section{Related Work}

A  range of formal models
 have been proposed to characterize the semantics of  isolation levels.
One influential formalism is \emph{dependency graphs},
  introduced by Bernstein et al.~\cite{bernstein1981concurrency} to characterize \ser
  and later extended by Adya~\cite{Adya:PhDThesis1999} to weaker isolation guarantees
   such as \si and \rc.
A second line of work adopts axiomatic semantics.
Cerone et al.~\cite{Framework:CONCUR2015} introduced 
\emph{abstract executions}
 that define isolation levels using two relations: visibility, capturing which effects a transaction may observe, and arbitration, ordering committed versions.
Biswas and Enea~\cite{Complexity:OOPSLA2019} later proposed an alternative axiomatization based on the write-read relation, which can be extracted directly from execution histories.
Orthogonal to these  models, 
Crooks et al.~\cite{Seeing:PODC2017} characterized isolation levels using 
\emph{state-based} traces, 
while Xiong et al.~\cite{Centralised:ECOOP2020} proposed an 
\emph{operational semantics} for them.
More recently, 
several alternative formalisms have been proposed specifically for black-box isolation checking. 
These include polygraphs and their variants~\cite{Cobra:OSDI2020,PolySI:VLDB2023,veristrong,Viper:EuroSys2023},
as well as transactional anomalous patterns~\cite{Plume:OOPSLA2024}.

Across these efforts, new formalizations are typically validated by proving their semantic equivalence to earlier ones.
  For example, Biswas and Enea~\cite{Complexity:OOPSLA2019} proved that 
  their axiomatic semantics is equivalent to that  of Cerone et al.~\cite{Framework:CONCUR2015}; Crooks et al.~\cite{Seeing:PODC2017} established the 
 correspondence between their state-based semantics and Adya's dependency-graph model~\cite{Adya:PhDThesis1999};
Huang et al.~\cite{PolySI:VLDB2023} proved that their characterization of \si using generalized polygraphs
 is equivalent to Cerone et al.'s dependency-graph-based characterization~\cite{AnalysingSI:JACM2018}.
However, all of these equivalence results rely solely on pen-and-paper proofs.
In contrast, we provide machine-checked proofs that  TAPs are sound and complete with respect to the isolation semantics they characterize.

Tool support for formal reasoning about database isolation 
  has also received increasing attention.
Liu et al.~\cite{ConsMaude:TACAS2019} developed a framework in  Maude~\cite{AllAboutMaude:Book2007} for specifying and model checking database isolation guarantees 
over execution histories of  transaction systems.
VerIso~\cite{VerIso:VLDB2025} mechanized the operational semantics of~\cite{Centralised:ECOOP2020} in Isabelle/HOL~\cite{DBLP:books/sp/NipkowPW02}, enabling machine-checked verification that concurrency control protocols satisfy their intended isolation guarantees. 
Our work differs from these efforts in that, 
  rather than verifying the correctness of protocols/systems, 
we mechanize the soundness and completeness proofs establishing the equivalence between  two  characterizations of  isolation levels.

More recently, Isolde~\cite{Isolde:arXiv2026} addresses the same goal as ours 
by automatically reasoning about the equivalence between different definitions of isolation levels.
Relying on bounded counterexample synthesis, 
it is sound
 (e.g., it correctly reports the \ra violation shown in Figure~\ref{fig:bug})
 and complete  within  a bound.
In contrast,
we establish 
fully machine-checked soundness and completeness proofs,
 at the cost of requiring interactive theorem proving.


\section{Conclusion}

We have presented a machine-checked formalization of TAPs and their associated HAT-compatible isolation levels in Rocq. 
We have also established machine-checked proofs of the corresponding soundness and completeness theorems. 
Beyond increasing confidence in the correctness of these results,
  our mechanization uncovered 
  an imprecise definition of transaction histories 
  and 
  subtle mismatches between TAP definitions and their intended semantics,
   leading to refinements that have since been incorporated into Plume.

We therefore argue that mechanization should not be viewed as an optional add-on,
but rather as 
a necessary step
in the rigorous formalization of isolation levels and consistency theories in general.
Existing formalizations can benefit from retrospective mechanization,
as this work illustrates, 
while future proposals such as~\cite{xiong2026semanticconformanceconcurrencycontrol,sccp-predicate} 
 should be accompanied by machine-checked developments from the outset.

Looking forward, 
we plan to develop a 
 meta-framework 
 for database isolation semantics, 
 which provides a common foundation for expressing diverse formalizations and establishing their semantic equivalence in a principled manner.

\section*{Acknowledgments}
Si Liu would like to express his deepest gratitude to David Basin for his generous support
and
 mentorship
  throughout his six years at ETH Zurich.
 He is especially grateful for the freedom to continue his research
 on the formal verification of distributed systems
 (e.g., \cite{10.1145/3563299,10.1145/3494517,fault-maude}),
 while also
 pursuing a new research direction on the verification and validation of database isolation guarantees
(e.g.,~\cite{PolySI:VLDB2023,TxCheck:OSDI2023,Plume:OOPSLA2024,VerIso:VLDB2025,MiniTransactions:ICDE2025,veristrong}).
He also sincerely thanks David for introducing him to the field of network security,
 particularly DNS security,
where they
and their collaborators
 made important advances
 (e.g., \cite{10.1145/3603269.3604870,298012,10.1145/3704898,DBLP:conf/csfw/WeghornLSPB22}).
Si Liu was supported by the University Development Fund (UDF01004607).
Hengfeng Wei was supported by
the Fundamental Research Funds for the Central Universities.

\bibliographystyle{splncs04}
\bibliography{ref}

@article{PolySI:VLDB2023,
	author = {Kaile Huang and Si Liu and Zhenge Chen and Hengfeng Wei and David A. Basin and Haixiang Li and Anqun Pan},
	title = {Efficient Black-box Checking of Snapshot Isolation in Databases},
	journal = {Proc. {VLDB} Endow.},
	volume = {16},
	number = {6},
	pages = {1264--1276},
	year = {2023},
	  doi          = {10.14778/3583140.3583145},
}

@inproceedings{Viper:EuroSys2023,
	author = {Zhang, Jian and Ji, Ye and Mu, Shuai and Tan, Cheng},
	title = {Viper: A Fast Snapshot Isolation Checker},
	year = {2023},
	publisher = {ACM},
	booktitle = {EuroSys 2023},
	pages = {654–671},
	doi = {10.1145/3552326.3567492},
}

@inproceedings{Cobra:OSDI2020,
	author       = {Cheng Tan and
	Changgeng Zhao and
	Shuai Mu and
	Michael Walfish},
	title        = {Cobra: Making Transactional Key-Value Stores Verifiably Serializable},
	booktitle    = {{OSDI} 2020},
	pages        = {63--80},
	publisher    = {{USENIX} Association},
	year         = {2020},
	  url          = {https://www.usenix.org/conference/osdi20/presentation/tan},
}

@article{Plume:OOPSLA2024,
	author       = {Si Liu and
	Long Gu and
	Hengfeng Wei and
	David Basin},
	title        = {Plume: Efficient and Complete Black-Box Checking of Weak Isolation
	Levels},
	journal      = {Proc. {ACM} Program. Lang.},
	volume       = {8},
	number       = {{OOPSLA}},
	pages        = {876--904},
	year         = {2024},
	doi          = {10.1145/3689742}
}

@article{Elle:VLDB2020,
	author = {Kingsbury, Kyle and Alvaro, Peter},
	title = {Elle: Inferring Isolation Anomalies from Experimental Observations},
	year = {2020},
	issue_date = {November 2020},
	publisher = {VLDB Endowment},
	volume = {14},
	number = {3},
	issn = {2150-8097},
	journal = {Proc. VLDB Endow.},
	pages = {268-280},
	numpages = {13},
  doi          = {10.5555/3430915.3442427},
}

@article{Complexity:OOPSLA2019,
	author       = {Ranadeep Biswas and
	Constantin Enea},
	title        = {On the complexity of checking transactional consistency},
	journal      = {Proc. {ACM} Program. Lang.},
	volume       = {3},
	number       = {{OOPSLA}},
	pages        = {165:1--165:28},
	year         = {2019},
	doi          = {10.1145/3360591}
}

@inproceedings {TxCheck:OSDI2023,
	author = {Zu-Ming Jiang and Si Liu and Manuel Rigger and Zhendong Su},
	title = {Detecting Transactional Bugs in Database Engines via {Graph-Based} Oracle Construction},
	booktitle = {OSDI 23},
	year = {2023},
	pages = {397--417},
	publisher = {USENIX Association}
}

@inproceedings{ConsMaude:TACAS2019,
	author       = {Si Liu and Peter Csaba {\"{O}}lveczky and Min Zhang and Qi Wang and Jos{\'{e}} Meseguer},
	title        = {Automatic Analysis of Consistency Properties of Distributed Transaction Systems in {Maude}},
	booktitle    = {{TACAS} 2019},
	series       = {LNCS},
	volume       = {11428},
	pages        = {40--57},
	publisher    = {Springer},
	year         = {2019},
	  doi          = {10.1007/978-3-030-17465-1\_3},
}

@article{critique,
	title={A critique of {ANSI} {SQL} isolation levels},
	author={Berenson, Hal and Bernstein, Phil and Gray, Jim and Melton, Jim and O'Neil, Elizabeth and O'Neil, Patrick},
	journal={ACM SIGMOD Record},
	volume={24},
	number={2},
	pages={1--10},
	year={1995},
	publisher={ACM},
	  doi          = {10.1145/223784.223785},
}

@phdthesis{Adya:PhDThesis1999,
	author = {Adya, Atul},
	title = {Weak Consistency: A Generalized Theory and Optimistic Implementations for Distributed Transactions},
	year = {1999},
	publisher = {Massachusetts Institute of Technology},
	address = {USA}
}

@article{AnalysingSI:JACM2018,
	author = {Cerone, Andrea and Gotsman, Alexey},
	title = {Analysing Snapshot Isolation},
	year = {2018},
	issue_date = {March 2018},
	publisher = {ACM},
	volume = {65},
	number = {2},
	journal = {J. ACM},
	month = jan,
	articleno = {11},
	  doi          = {10.1145/3152396},
}

@inproceedings{Framework:CONCUR2015,
  author       = {Andrea Cerone and
Giovanni Bernardi and
Alexey Gotsman},
title        = {A Framework for Transactional Consistency Models with Atomic Visibility},
booktitle    = {{CONCUR} 2015},
series       = {LIPIcs},
volume       = {42},
pages        = {58--71},
publisher    = {Schloss Dagstuhl - Leibniz-Zentrum f{\"{u}}r Informatik},
year         = {2015},
  doi          = {10.4230/LIPICS.CONCUR.2015.58},
}

@article{veristrong,
	author = {Cai, Zhiheng and Liu, Si and Wei, Hengfeng and Chen, Yuxing and Pan, Anqun},
	title = {Fast Verification of Strong Database Isolation},
	year = {2025},
	publisher = {VLDB Endowment},
	volume = {19},
	number = {4},
	issn = {2150-8097},
	doi = {10.14778/3785297.3785300},
	journal = {Proc. VLDB Endow.},
	pages = {563–575},
}

@inproceedings{Seeing:PODC2017,
	author = {Crooks, Natacha and Pu, Youer and Alvisi, Lorenzo and Clement, Allen},
	title = {Seeing is Believing: A Client-Centric Specification of Database Isolation},
	year = {2017},
	isbn = {9781450349925},
	publisher = {ACM},
	booktitle = {Proceedings of the ACM Symposium on Principles of Distributed Computing},
	pages = {73–82},
	numpages = {10},
	series = {PODC '17},
    doi = {10.1145/3087801.3087802},
}

@article{RAMP:TODS2016,
	author = {Bailis, Peter and Fekete, Alan and Ghodsi, Ali and Hellerstein, Joseph M. and Stoica, Ion},
	title = {Scalable Atomic Visibility with {RAMP} Transactions},
	year = {2016},
	publisher = {ACM},
	volume = {41},
	number = {3},
	journal = {ACM Trans. Database Syst.},
	month = jul,
	articleno = {15},
	  doi          = {10.1145/2909870},
}

@inproceedings{Centralised:ECOOP2020,
	author       = {Shale Xiong and
	Andrea Cerone and
	Azalea Raad and
	Philippa Gardner},
	title        = {Data Consistency in Transactional Storage Systems: {A} Centralised
	Semantics},
	booktitle    = { {ECOOP} 2020},
	series       = {LIPIcs},
	volume       = {166},
	pages        = {21:1--21:31},
	publisher    = {Schloss Dagstuhl - Leibniz-Zentrum f{\"{u}}r Informatik},
	year         = {2020},
	  doi          = {10.4230/LIPICS.ECOOP.2020.21},
}

@article{hat,
	author = {Bailis, Peter and Davidson, Aaron and Fekete, Alan and Ghodsi, Ali and Hellerstein, Joseph M. and Stoica, Ion},
	title = {Highly Available Transactions: Virtues and Limitations},
	year = {2013},
	issue_date = {November 2013},
	publisher = {VLDB Endowment},
	volume = {7},
	number = {3},
	journal = {Proc. VLDB Endow.},
	month = {nov},
	pages = {181–192},
	numpages = {12},
doi          = {10.14778/2732232.2732237},
}

@inproceedings{MiniTransactions:ICDE2025,
	author       = {Hengfeng Wei and
	Jiang Xiao and
	Na Yang and
	Si Liu and
	Zijing Yin and
	Yuxing Chen and
	Anqun Pan},
	title        = {Boosting End‑to‑End Database Isolation Checking via Mini‑Transactions},
	booktitle    = {ICDE 2025},
	year = {2025},
pages = {3998-4010},
publisher = {IEEE Computer Society},
month =May,
  doi          = {10.1109/ICDE65448.2025.00298},
}

@article{VerIso:VLDB2025,
	title={{VerIso:} Verifiable Isolation Guarantees for Database Transactions},
	author={Shabnam Ghasemirad and Si Liu and Christoph Sprenger and Luca Multazzu and David Basin},
	journal      = {Proc. {VLDB} Endow.},
	volume       = {18},
	number       = {5},
	pages        = {1362--1375},
	year         = {2025},
	  doi          = {10.14778/3718057.3718065},
}

@article{AWDIT:PLDI2025,
author = {M\o{}ldrup, Lasse and Pavlogiannis, Andreas},
title = {{AWDIT:} An Optimal Weak Database Isolation Tester},
year = {2025},
issue_date = {June 2025},
publisher = {ACM},
volume = {9},
number = {PLDI},
journal = {Proc. ACM Program. Lang.},
month = jun,
articleno = {209},
numpages = {25},
doi = {10.1145/3742465},
}

@book{TIS:Book2002,
	author       = {Gerhard Weikum and
	Gottfried Vossen},
	title        = {Transactional Information Systems: Theory, Algorithms, and the Practice
	of Concurrency Control and Recovery},
	publisher    = {Morgan Kaufmann},
	year         = {2002},
	isbn         = {1-55860-508-8}
}

@article{IsoVista:VLDB24,
	author = {Gu, Long and Liu, Si and Xing, Tiancheng and Wei, Hengfeng and Chen, Yuxing and Basin, David},
	title = {{IsoVista:} Black-Box Checking Database Isolation Guarantees},
	year = {2024},
	issue_date = {August 2024},
	publisher = {VLDB Endowment},
	volume = {17},
	number = {12},
	journal = {Proc. VLDB Endow.},
	month = aug,
	pages = {4325–4328},
	numpages = {4},
	doi = {10.14778/3685800.3685866},
}

@article{noc-noc,
	author = {Liu, Si and Multazzu, Luca and Wei, Hengfeng and Basin, David A.},
	title = {{NOC-NOC:} Towards Performance-optimal Distributed Transactions},
	year = {2024},
	issue_date = {February 2024},
	publisher = {ACM},
	volume = {2},
	number = {1},
	journal = {Proc. ACM Manag. Data},
	month = mar,
	articleno = {9},
	numpages = {25},
	doi = {10.1145/3639264},
}

@misc{rocq,
	title={{The Rocq Prover}}, 
	author={Rocq},
	year={Accessed in June, 2026},
	url={https://rocq-prover.org/}, 
}

@inproceedings{Cure:ICDCS2016,
	author    = {Deepthi Devaki Akkoorath and
	Alejandro Z. Tomsic and
	Manuel Bravo and
	Zhongmiao Li and
	Tyler Crain and
	Annette Bieniusa and
	Nuno M. Pregui{\c{c}}a and
	Marc Shapiro},
	title     = {Cure: Strong Semantics Meets High Availability and Low Latency},
	booktitle = { {ICDCS} 2016},
	pages     = {405--414},
	publisher = {{IEEE} Computer Society},
	year      = {2016},
	  doi          = {10.1109/ICDCS.2016.98},
}

@inproceedings{NOC:OSDI2020,
	author = {Haonan Lu and Siddhartha Sen and Wyatt Lloyd},
	title = {{Performance-Optimal} {Read-Only} Transactions},
	booktitle = {14th USENIX Symposium on Operating Systems Design and Implementation (OSDI 20)},
	year = {2020},
	isbn = {978-1-939133-19-9},
	pages = {333--349},
	publisher = {USENIX Association},
	month = nov,
 	url = {https://www.usenix.org/conference/osdi20/presentation/lu},
}

@inproceedings{9139787,
	author={Antoniadis, Karolos and Didona, Diego and Guerraoui, Rachid and Zwaenepoel, Willy},
	booktitle={2020 IEEE International Parallel and Distributed Processing Symposium (IPDPS)}, 
	title={The Impossibility of Fast Transactions}, 
	year={2020},
	volume={},
	number={},
	pages={1143-1154},
	doi={10.1109/IPDPS47924.2020.00120}
}

@article{bernstein1981concurrency,
	title={Concurrency control in distributed database systems},
	author={Bernstein, Philip A and Goodman, Nathan},
	journal={ACM Computing Surveys (CSUR)},
	volume={13},
	number={2},
	pages={185--221},
	year={1981},
	publisher={ACM},
	doi          = {10.1145/356842.356846}
}

@book{AllAboutMaude:Book2007,
	author = {Clavel, Manuel and Dur\'{a}n, Francisco and Eker, Steven and Lincoln, Patrick and Mart\'{\i}-Oliet, Narciso and Meseguer, Jos\'{e} and Talcott, Carolyn},
	title = {All about Maude - a High-Performance Logical Framework: How to Specify, Program and Verify Systems in Rewriting Logic},
	year = {2007},
	isbn = {3540719407},
	publisher = {Springer-Verlag},
	address = {Berlin, Heidelberg},
	  doi          = {10.1007/978-3-540-71999-1},
}

@book{DBLP:books/sp/NipkowPW02,
	author = {Tobias Nipkow and Lawrence C. Paulson and Markus Wenzel},
	doi = {10.1007/3-540-45949-9},
	isbn = {3-540-43376-7},
	publisher = {Springer},
	series = {Lecture Notes in Computer Science},
	title = {Isabelle/HOL - {A} Proof Assistant for Higher-Order Logic},
	volume = {2283},
	year = {2002}
}

@misc{Isolde:arXiv2026,
      title={Reasoning about Transactional Isolation Levels with Isolde}, 
      author={Manuel Barros and Alcino Cunha and Jose Pereira and Eunsuk Kang},
      year={2026},
      eprint={2604.00159},
      archivePrefix={arXiv},
      primaryClass={cs.DB},
      url={https://arxiv.org/abs/2604.00159}, 
}

@inproceedings{10.1145/3603269.3604870,
	author = {Liu, Si and Duan, Huayi and Heimes, Lukas and Bearzi, Marco and Vieli, Jodok and Basin, David and Perrig, Adrian},
	title = {{A Formal Framework for End-to-End DNS Resolution}},
	year = {2023},
	isbn = {9798400702365},
	publisher = {ACM},
	doi = {10.1145/3603269.3604870},
	booktitle = {Proceedings of the ACM SIGCOMM 2023 Conference},
	pages = {932–949},
	numpages = {18},
	series = {ACM SIGCOMM '23}
}

@inproceedings {298012,
	author = {Huayi Duan and Marco Bearzi and Jodok Vieli and David Basin and Adrian Perrig and Si Liu and Bernhard Tellenbach},
	title = {{CAMP}: Compositional Amplification Attacks against {DNS}},
	booktitle = {USENIX Security 24},
	year = {2024},
	isbn = {978-1-939133-44-1},
	address = {Philadelphia, PA},
	pages = {5769--5786},
	url = {https://www.usenix.org/conference/usenixsecurity24/presentation/duan},
	publisher = {USENIX Association},
	month = aug
}

@article{10.1145/3704898,
	author = {Nevatia, Dhruv and Liu, Si and Basin, David},
	title = {{Reachability Analysis of the Domain Name System}},
	year = {2025},
	issue_date = {January 2025},
	publisher = {ACM},
	volume = {9},
	number = {POPL},
	doi = {10.1145/3704898},
	journal = {Proc. ACM Program. Lang.},
	month = jan,
	articleno = {62},
	numpages = {31}
}

@inproceedings{DBLP:conf/csfw/WeghornLSPB22,
	author       = {Thilo Weghorn and
	Si Liu and
	Christoph Sprenger and
	Adrian Perrig and
	David A. Basin},
	title        = {{N-Tube: Formally Verified Secure Bandwidth Reservation in Path-Aware
	Internet Architectures}},
	booktitle    = {{CSF} '22},
	pages        = {147--162},
	publisher    = {{IEEE}},
	year         = {2022},
	doi          = {10.1109/CSF54842.2022.9919646},
}

@article{DBLP:journals/cacm/Terry13,
	author       = {Doug Terry},
	title        = {Replicated data consistency explained through baseball},
	journal      = {Commun. {ACM}},
	volume       = {56},
	number       = {12},
	pages        = {82--89},
	year         = {2013},
	doi          = {10.1145/2500500},
}

@article{10.1145/3494517,
	author = {Liu, Si},
	title = {{All in One: Design, Verification, and Implementation of SNOW-optimal Read Atomic Transactions}},
	year = {2022},
	issue_date = {July 2022},
	publisher = {ACM},
	volume = {31},
	number = {3},
	issn = {1049-331X},
	doi = {10.1145/3494517},
	journal = {ACM Trans. Softw. Eng. Methodol.},
	month = mar,
	articleno = {43},
	numpages = {44}
}

@inproceedings{EPplus,
	author    = {Shabnam Ghasemirad and Christoph Sprenger and Si Liu and Luca Multazzu and David Basin},
	title     = {Pushing the Limit: Verified Performance-Optimal Causally-Consistent Database Transactions},
	booktitle = {TACAS '25},
	series    = {LNCS},
	volume    = {15698},
	pages     = {43--62},
	year      = {2025},
	doi       = {10.1007/978-3-031-90660-2\_3},
}

@article{10.1145/3563299,
	author = {Liu, Si and Meseguer, Jose and \"{O}lveczky, Peter Csaba and Zhang, Min and Basin, David},
	title = {Bridging the semantic gap between qualitative and quantitative models of distributed systems},
	year = {2022},
	issue_date = {October 2022},
	publisher = {ACM},
	volume = {6},
	number = {OOPSLA2},
	doi = {10.1145/3563299},
	journal = {Proc. ACM Program. Lang.},
	month = oct,
	articleno = {136},
	numpages = {30}
}

@inproceedings{DBLP:conf/tase/LiangL21,
  author       = {Lei Liang and
                  Si Liu},
  title        = {Exploring Design Alternatives for Replicated {RAMP} Transactions Using
                  Maude},
  booktitle    = {{TASE} 2021},
  pages        = {111--118},
  publisher    = {{IEEE}},
  year         = {2021},
  doi          = {10.1109/TASE52547.2021.00026}
}

@inproceedings{fault-maude,
	author = {Zhou, Ziwei and Liu, Si and Zhou, Zhou and Wang, Peixin and Zhang, Min},
	title = {A Formal Framework for Predicting Distributed System Performance Under Faults},
	year = {2026},
	isbn = {978-3-032-26203-5},
	publisher = {Springer-Verlag},
	doi = {10.1007/978-3-032-26204-2_3},
	booktitle = {Formal Methods: 27th International Symposium, FM 2026},
	pages = {47–66},
	numpages = {20},
}

@misc{xiong2026semanticconformanceconcurrencycontrol,
      title={Semantic Conformance of Concurrency Control Protocols under Mixed Isolation Levels}, 
      author={Qiuhuan Xiong and Hengfeng Wei and Si Liu and Yuxing Chen and Jidong Ge},
      year={2026},
      eprint={2607.16696},
      archivePrefix={arXiv},
      primaryClass={cs.DB},
      url={https://arxiv.org/abs/2607.16696}, 
}

@misc{sccp-predicate,
      title={Specifying and Analyzing Transactional Consistency Models with
Predicates (Extended Abstract)}, 
      author={Hengfeng Wei and Si Liu and Yuxing Chen},
      year={2026},
      url={https://sccp-workshop.github.io/sccp26/papers/sccp26-consistency-models-predicates.pdf}, 
      note={presented at SCCP 2026}
}


\end{document}